\documentclass[12pt]{article}

\usepackage{amsmath}
\begin{document}
\setcounter{page}{1}
\def\theequation{\arabic{section}.\arabic{equation}} 
\newcommand{\be}{\begin{equation}}
\newcommand{\ee}{\end{equation}}
\newcommand{\ul}{\underline}
\begin{titlepage}
\title{A symmetry of the Einstein-Friedmann equations for spatially flat, 
perfect fluid, universes}

\author{Valerio Faraoni \\ 
       \\ {\small Physics Department, Bishop's University} \\
{\small 2600 College Street, Sherbrooke, Qu\'ebec, Canada J1M 1Z7}\\
{\small  vfaraoni@ubishops.ca}
}
\date{} 
\maketitle
\thispagestyle{empty}

\begin{abstract}

We report a symmetry property of the Einstein-Friedmann equations for 
spatially flat Friedmann-Lema\^{i}tre-Robertson-Walker universes filled 
with a perfect fluid with any constant equation of state. The symmetry 
transformations form a one-parameter Abelian group.

\end{abstract}  
\end{titlepage}   \clearpage

\section{Introduction}

In the recent literature, there have been many studies of the symmetries 
of 
the Einstein-Friedmann equations of spatially homogeneous and isotropic 
Friedmann-Lema\^itre-Robertson-Walker (FLRW) cosmology. These studies are 
ultimately inspired by (although not always directy related to) string 
dualities \cite{etc1, etc2, etc3, etc4, etc5, etc6, etc7, etc8, etc9, 
etc10, etc11, etc12, etc13, etc14, etc15, etc16, etc17, etc18, etc19, 
etc20, myPLB} or by methods introduced in supersymmetric quantum mechanics 
\cite{etc21, etc22, etc23}. Here we discuss a simpler symmetry arising in 
the context of pure Einstein gravity.  Adopting  
the notation of Refs.~\cite{Wald, Carroll}, we use metric signature 
$-+++$ and units in which the speed of light and Newton's constant are 
unity.

The FLRW line element of spatially homogeneous and isotropic cosmology in 
comoving coordinates$\left( t,x,y,z \right)$ is
\be
ds^2 =-dt^2 +a^2(t) \left( dx^2+dy^2 +dz^2 \right)\,,
\ee
where the dynamics is contained in the evolution of the cosmic scale 
factor $a(t)$.
The Einstein-Friedmann equations for a spatially flat universe filled with 
a perfect fluid with energy density $\rho(t)$ and isotropic pressure 
$P(t)$ are
\begin{eqnarray}
&& H^2 = \frac{8\pi}{3} \, \rho \,, \label{eq:general1}\\
&&\nonumber\\
&& \dot{\rho}+3H\left( P+\rho \right) =0 \,,\label{eq:general2}\\
&&\nonumber\\
&& \frac{ \ddot{a}}{a} = -\, \frac{4\pi}{3} \left( \rho+3P \right) 
\,.\label{eq:general3}
\end{eqnarray}
These equations exhibit a special symmetry that maps a  
barotropic perfect fluid 
with equation of state $P=P(\rho)$ into itself, with a rescaled 
energy density and pressure but with the same equation of state.

\section{The symmetry transformation} 

Assume that the energy content of the universe is a single perfect fluid 
with 
constant equation of state $P=w\rho$ with $w=$~constant;  then the 
Einstein-Friedmann equations~(\ref{eq:general1})-(\ref{eq:general3}) 
assume the form
\begin{eqnarray}
&& H^2 = \frac{8\pi}{3} \, \rho \,, \label{EF1}\\
&&\nonumber\\
&& \dot{\rho} + 3\left( w+1 \right) \frac{\dot{a}}{a} \, \rho =0 
\,,\label{EF2}\\
&&\nonumber\\
&& \frac{ \ddot{a}}{a} = -\, \frac{4\pi}{3} \left( 3w+1 \right) \rho
\,.\label{EF3}
\end{eqnarray}
The solution of these equations is well known and easy to 
derive ({\em e.g.}, \cite{Wald}): 
\begin{eqnarray}
a(t) &=& a_0 \, t^{ \frac{2}{3(w+1)} } \,,\label{a-solution}\\
&&\nonumber\\
\rho(a) &=& \frac{ \rho_0}{ a^{3(w+1)} } \,. \label{rho-solution}
\end{eqnarray}
The change of variables  
\begin{eqnarray}
a & \longrightarrow & \tilde{a} = a^s \,,\label{change1}\\
&& \nonumber\\
dt & \longrightarrow & d\tilde{t} = s \, a^{ 
\frac{3(w+1)(s-1)}{2} } dt = s \, \tilde{a}^{ \frac{3(w+1)(s-1)}{2s} }dt   
\,, \label{change2}\\ 
&& \nonumber\\
\rho & \longrightarrow & \tilde{\rho} = a^{-3(w+1)(s-1)} \rho 
\,,\label{change3}
\end{eqnarray}
with inverse
\begin{eqnarray}
a & = & \tilde{a}^{1/s} \,,\\
&& \nonumber\\
dt & = &  \frac{ a^{\frac{ -3(w+1)(s-1)}{s}} }{s} \,  d\tilde{t} \,,\\
&& \nonumber\\
\rho & =  & \tilde{\rho} \, \tilde{a}^{ \frac{ 3(w+1)(s-1)}{s} } \,,
\end{eqnarray}
leaves the Einstein-Friedmann equations~(\ref{EF1})-(\ref{EF3}) 
unchanged. In particular, the matter source, the barotropic perfect fluid, 
maintains 
the equation of state $P=w\rho$ with the same equation of state parameter 
$w$. This is in contrast with other symmetries of the same equations which 
change an ``ordinary'' fluid satisfying the weak energy condition into a 
phantom fluid with different equation of state parameter and are 
ultimately inspired by string theory dualities 
\cite{myPLB, etc1, etc2, etc3, etc4, etc5, etc6, etc7, etc8, etc9, 
etc10, etc11, etc12, etc13, etc14, etc15, etc16, etc17, etc18, etc19, 
etc20}, and with other solutions of the Einstein–Friedmann 
equations obtained  using  methods of
supersymmetric quantum mechanics \cite{etc21, etc22, etc23}.

To check invariance of the equations, begin from the Friedmann equation 
(\ref{EF1}) which becomes, in terms of tilded quantities:
\be
\frac{1}{ \tilde{a}^{2/s}} \left[ \frac{ d\left( \tilde{a}^{1/s} 
\right)}{d\tilde{t} } \right]^2 \left( \frac{ d\tilde{t}}{dt} \right)^2 
=\frac{8\pi }{3} \, \tilde{\rho} \, \tilde{a}^{ \frac{3(w+1)(s-1)}{s}} 
\ee
or 
\be
\frac{1}{s^2} \, \frac{ \tilde{a}^{ \frac{2}{s}-2} }{ \tilde{a}^{2/s} }
\left( \frac{ d \tilde{a} }{d\tilde{t} } \right)^2 s^2  \tilde{a}^{ 
\frac{3(w+1)(s-1)}{s} } = \frac{8\pi }{3} \, \tilde{\rho}\tilde{a}^{
\frac{3(w+1)(s-1)}{s} } \,,
\ee
which simplifies to 
\be
\left( \frac{1}{\tilde{a}} \, \frac{ d\tilde{a}}{d\tilde{t}} \right)^2 
= \frac{8\pi }{3} \, \tilde{\rho} \,.
\ee

Let us consider now the covariant conservation equation~(\ref{EF2}), which 
now reads
\be
\frac{d}{d\tilde{t}} \left( \tilde{\rho} \, 
\tilde{a}^{\frac{3(w+1)(s-1)}{s}} \right) \frac{d \tilde{t}}{dt} 
+3(w+1)\tilde{\rho} \,\, \frac{ \tilde{a}^{\frac{3(w+1)(s-1)}{s}} }{ 
\tilde{a}^{1/s} }\, \frac{d \tilde{t}}{dt} \,  \frac{ d\left( 
\tilde{a}^{1/s}\right) }{d\tilde{t}} =0 \,.
\ee
Expanding, one has
\be
\tilde{a}^{ \frac{3(w+1)(s-1)}{s} } \, \frac{ d\tilde{\rho} }{d \tilde{t}} 
+\frac{3(w+1)(s-1)}{s} \, \tilde{a}^{ \frac{3(w+1)(s-1)}{s} -1} \, 
\frac{ d\tilde{a}}{d\tilde{t}}   \, \tilde{\rho} +\frac{3(w+1)}{s} \, 
\tilde{a}^{ \frac{3(w+1)(s-1)-1}{s}} \, \tilde{a}^{\frac{1}{s}-1} \, 
\tilde{\rho} \, 
\frac{ d\tilde{a}}{d\tilde{t}}=0 
\,.
\ee
Grouping similar terms then yields 
\be
\tilde{a}^{ \frac{3(w+1)(s-1)}{s} } \left[ \frac{ d\tilde{\rho}}{d 
\tilde{t}} +3\left(w+1\right) \frac{1}{ \tilde{a}} \, \frac{ 
d\tilde{a}}{d\tilde{t}}  \, \tilde{\rho} \right]=0 \,.
\ee
Since $\tilde{a} \neq 0$, the conservation equation in the tilded world 
follows.

Coming to the acceleration equation (\ref{EF3}), one first notices that 
\be
\frac{da}{dt}= \frac{ d\left( \tilde{a}^{1/s}\right)}{d \tilde{t}} 
\, \frac{ d \tilde{t}}{dt} = 
\tilde{a}^{\frac{(3w+1)(s-1)}{2s} } \, \frac{d\tilde{a}}{d\tilde{t}} \,,
\ee

\be
\frac{d^2a}{dt^2}= \frac{ d \tilde{t}}{dt} \, \frac{d}{d\tilde{t}} 
\left( \frac{da}{dt} \right)= \tilde{a}^{\frac{(3w+1)(s-1)}{s} } \left[ 
\frac{(3w+1)(s-1)}{2} \, \tilde{a} \left( \frac{1}{ \tilde{a}} \, 
\frac{d\tilde{a} }{ d\tilde{t}}  
\right)^2 
+s \, \frac{ d^2  \tilde{a}}{d \tilde{t}^2} \right] \,.
\ee
Using the Friedmann equation for tilded quantities, this becomes 
\be
\frac{d^2a}{dt^2}= 
\tilde{a}^{ \frac{(3w+1)(s-1)}{s}+1} \left[ \frac{(3w+1)(s-1)}{2}\, 
\frac{8\pi}{3} \, \tilde{\rho} +\frac{s}{ \tilde{a}} \,  \frac{ d^2  
\tilde{a}}{d \tilde{t}^2} \right] \,.
\ee
The acceleration equation~(\ref{EF3}) now reads
\be
\frac{ \tilde{a}^{ \frac{(3w+1)(s-1)}{s}+1}}{ \tilde{a}^{ 1/s}}\left[ 
(3w+1)(s-1) \frac{4\pi}{3} \, \tilde{\rho} +\frac{s}{ \tilde{a}} \,
\frac{ d^2 \tilde{a}}{d \tilde{t}^2} \right] =-\frac{4\pi}{3} 
\left( 3w+1\right) \tilde{\rho} \, \tilde{a}^{ \frac{(3w+1)(s-1)}{s}} \,.
\ee
By simplifying a factor $\tilde{a}^{ \frac{(3w+1)(s-1)}{s}}$ on both sides 
and collecting similar terms, one obtains 
\be
\frac{s}{ \tilde{a}} \, \frac{ d^2 \tilde{a}}{d \tilde{t}^2}= 
-\frac{4\pi}{3} \left(3w+1\right) \tilde{\rho}\left[1+(s-1)\right]
\ee
and the acceleration equation for tilded quantities follows.

\section{A group of symmetry transformations}

The transformations (\ref{change1})-(\ref{change3}) form a commutative 
group, as shown below. First, we show that the composition of 
two such transformations is a change of variables of the same form. Let 
\begin{eqnarray}
\hat{L}_s : \;\;\; \left( a, dt, \rho \right) &\longrightarrow & 
\left( \tilde{a} , d\tilde{t}, \tilde{\rho} \right) =
\left( a^s,  s \, a^{\frac{3(w+1)(s-1)}{2} }  dt, a^{-3(w+1)(s-1)} \, \rho  
\right)   \,,\\
&&\nonumber\\
\hat{L}_p : \;\;\; \left( a, dt, \rho \right) & \longrightarrow & 
\left( \tilde{\tilde{a} } , d\tilde{\tilde{t} },\tilde{  \tilde{\rho}} 
\right) =
\left( a^p,  p\, a^{\frac{ 3(w+1)(p-1)}{2}}  dt, a^{-3(w+1)(p-1)} \, \rho 
\right)  \,;
\end{eqnarray}
then the composition of the two transformation gives
\be
a \rightarrow \tilde{a} \rightarrow \tilde{ \tilde{a} }=\tilde{a}^p = 
a^{ps} \equiv a^r \,,
\ee

\begin{eqnarray}
dt \rightarrow d\tilde{t} \rightarrow d\tilde{ \tilde{t} }&=& 
p \tilde{a}^{ \frac{3(w+1)(p-1)}{2} } d\tilde{t} = p a^{ 
\frac{3(w+1)(p-1)s }{2} } s a^{\frac{3(w+1)(s-1)}{2} } dt = 
ps a^{ \frac{3(w+1)(ps-1)}{2}} dt \nonumber\\
&&\nonumber\\
& \equiv & r a^{ 
\frac{3(w+1)(r-1)}{2}} dt\,,
\end{eqnarray}

\be
\rho \rightarrow \tilde{\rho} \rightarrow \tilde{ \tilde{\rho}} =
\tilde{a}^{-3(w+1)(p-1)} \, \tilde{\rho} = a^{-3(w+1)(p-1)s} 
a^{-3(w+1)(s-1)} \rho = a^{-3(w+1)(r-1)} \rho \,,
\ee
where $r =ps$. Therefore, the composition of two transformations gives 
the same kind of transformation and the order of these two operations does 
not matter:
\be
\hat{L}_s \circ \hat{L}_p = \hat{L}_p \circ \hat{L}_s \equiv \hat{L}_{sp} 
\equiv \hat{L}_r \,.
\ee

There is a neutral element for the operation of composition of maps: the 
transformation $\hat{L}_1 $ with $s=1$ is 
the identity since
\be 
\hat{L}_1 : \;\;\; \left( a, dt, \rho \right) \longrightarrow 
\left( \tilde{a} , d\tilde{t}, \tilde{\rho} \right) =  
\left( a, dt, \rho \right) \,.
\ee
Finally, each transformation $ \hat{L}_s $ with $s\neq 0$ has a (left and 
right) inverse $ \hat{L}_{1/s} $ since 
\be
\hat{L}_s \circ \hat{L}_{1/s} = \hat{L}_{1/s} \circ \hat{L}_s =
\hat{L}_{ \left( s \cdot 1/s \right) }= \hat{I}_d  \,.
\ee
Therefore, the transformations $\hat{L}_s$ given by 
Eqs.~(\ref{change1})-(\ref{change3}) form a one-parameter  Abelian group 
parametrized by the real number $s\neq 0$.

\section{Symmetry of the solutions}

Since the form of the Einstein-Friedmann equations (\ref{EF1})-(\ref{EF3}) 
does not change under the symmetry operation 
(\ref{change1})-(\ref{change3}), the solution corresponding to the 
same perfect fluid will still be given by Eqs.~(\ref{a-solution}) and 
(\ref{rho-solution}) but now with tilded scale factor:
\begin{eqnarray}
\tilde{a}(t) &=& a_0 \, \tilde{t}^{ \frac{2}{3(w+1)} } 
\,,\\
&&\nonumber\\
\tilde{\rho} ( \tilde{a}) &=& \frac{ \rho_0}{ \tilde{a}^{3(w+1)} } \,. 
\end{eqnarray}
Let us verify this property explicitly. We have, using 
Eq.~(\ref{a-solution}),
\be
d\tilde{t}= s\, a^{\frac{3(w+1)(s-1)}{2}} \, dt = 
s \, a_0^{\frac{3(w+1)(s-1)}{2}} \, t^{s-1} dt 
\ee
and, integrating, 
\be
\tilde{t}= \int d\tilde{t} = a_0^{\frac{3(w+1)(s-1)}{2}} t^s \,, 
\label{questa}
\ee
where we set to zero an additive integration constant assuming that 
$\tilde{t}(0)=0$ if $s>0$ (apart from dimensions, the multiplicative 
constant in the right hand side of Eq.~(\ref{questa}) is not physically 
relevant because it can be absorbed into a rescaling of the time 
coordinate). Inverting this 
relation yields  \be
t= \frac{\tilde{t}^{ 1/s}}{ a_0^{\frac{3(w+1)(s-1)}{2}} } \,.
\ee
Therefore, it is
\be
\tilde{a}( \tilde{t}) =  a^s = a_0^s \, t^{ \frac{ 2s}{3(w+1)}} = 
a_0^{\frac{2s^2 -3(w+1)s +3(w+1))}{2s}} \, \tilde{t}^{ \frac{2}{3(w+1)}} 
\equiv 
\tilde{a}_0 \, \tilde{t}^{ \frac{2}{3(w+1)}} \,.
\ee
Similarly, one checks that
\be
\tilde{\rho} ( \tilde{a}) = a^{ 3(w+1)(s-1)} \rho =  a^{ 3(w+1)(s-1)} \, 
\frac{\rho_0}{ 
a^{3(w+1)}} = \frac{ \rho_0}{ \tilde{a}^{ 3(w+1)} } \,.
\ee

\section{Conclusions}

We have reported a symmetry property of the Einstein-Friedmann equations 
for a spatially flat FLRW universe filled with a single barotropic perfect 
fluid with constant equation of state $P=w\rho$. It is easy to check that 
this symmetry does not hold for spatially curved universes.

As for the physical meaning of this symmetry, let us note that the scale 
factor and the comoving time scale as
\begin{eqnarray}
a &\rightarrow  &a^s \,,\\
&&\nonumber\\
t &\rightarrow  & t^s \,,
\end{eqnarray}
respectively. 
This scaling could be superficially interpreted by saying that the scaling 
of proper 
spatial distances and the proper time of observers comoving with the cosmic 
perfect fluid by the same power points to some 
scale-invariance property of the Einstein-Friedmann equations for 
spatially flat sections, with this property failing when a spatial scale 
associated 
with the curvature of the 3-dimensional spatial sections is present. 
However, 
these equations are definitely {\em not} scale invariant, and this is the 
meaning of the completely different scaling of the energy 
density~(\ref{change3}) (in the units used, in which energy and mass have 
the dimensions of a length, energy density should scale as the inverse 
square of a length $\ell^{-2s}$, but it does not). This fact simply 
reflects the lack of scale invariance of the Einstein equations even {\em 
in vacuo} or in the presence of conformally invariant matter (such as, 
{\em e.g.}, a radiation fluid with $w=1/3$).

Next, one could be tempted to view the symmetry of the Einstein-Friedmann 
equations (\ref{EF1})-(\ref{EF3}) as deriving from a conformal 
transformation of the spacetime metric $ds^2\rightarrow d\tilde{s}^2= 
\Omega^2 ds^2$ for some conformal factor $\Omega (x^{\mu})$, followed by a 
suitable redefinition of the comoving time coordinate, but this is not 
possible in general, as is easy to check.

The symmetry map (\ref{change1})-(\ref{change3}) for the special case of a 
radiation fluid with equation of state parameter $w=1/3$ was already noted 
in the context of an analogy between the cosmic radiation era and the 
freezing of bodies of water in environmental physics 
\cite{Faraoniunpublished}.

\section*{Acknowledgments} 

This work is supported by the Natural Sciences and Engineering Research 
Council of Canada (Grant No. 2016-03803) and by Bishop's University.


{\small \begin{thebibliography}{99}


\bibitem{etc1} Chimento, L.P. Symmetry and inflation {\em Phys. Rev. D} 
{\bf 2002}, 65, 063517. DOI: 10.1103/PhysRevD.65.063517

\bibitem{etc2} Aguirregabiria, J.M., Chimento, L.P., Jakubi, A., Lazkoz, 
R. Symmetries leading to inflation {\em Phys. Rev. D} {\bf 2003} 67, 
083518, DOI: 10.1103/PhysRevD.67.083518

\bibitem{etc3} Chimento, L.P., Lazkoz, R. On the link between phantom and 
standard cosmologies.  {\em Phys. Rev. Lett.} {\bf 2003}, 91, 211301, DOI: 
10.1103/PhysRevLett.91.211301

\bibitem{etc4} Dabrowski, M.P., Stachowiak, T., Szydlowski, M.  Phantom 
cosmologies  {\em Phys. Rev. D} {\bf 2003}, 68, 103519, DOI: 
10.1103/PhysRevD.68.103519

\bibitem{etc5} Aguirragabiria, J.M., Chimento, L.P., Lazkoz, R. Phantom 
k-essence cosmologies. {\em Phys. Rev. D} {\bf 2004}, 70, 023509, DOI: 
10.1103/PhysRevD.70.023509

\bibitem{etc6} Calcagni, G. Patch dualities and remarks on nonstandard 
cosmologies. {\em Phys. Rev. D} {\bf 2005}, 71, 023511, DOI: 
10.1103/PhysRevD.71.023511

\bibitem{etc7} Szydlowski, M., Godlowski, W., Wojtak, R. Equation of state 
for Universe from similarity symmetries. {\em Gen. Relativ. Gravit.} {\bf 
2006}, 38, 795, DOI: 10.1007/s10714-006-0265-6

\bibitem{etc8} Chimento, L.P., Lazkoz, R. Duality extended Chaplygin 
cosmologies with a big rip. {\em Class. Quantum Grav.} {\bf 2006}, 23, 
3195-3204, DOI: 10.1088/0264-9381/23/9/027

\bibitem{etc9} Chimento, L.P., Zimdhal, W. Duality invariance and 
cosmological dynamics. {\em Int. J. Mod. Phys. D} {\bf 2008}, 17, 
2229-2254, DOI: 10.1142/S0218271808013820

\bibitem{etc10} Chimento, L.P., Pavon, D. Dual interacting cosmologies and 
late accelerated expansion. {\em Phys. Rev. D} {\bf 2006}, 73, 063511, 
DOI: 10.1103/PhysRevD.73.063511

\bibitem{etc11} Dabrowski, M.P., Kiefer, C., Sandhoefer, B. Duality 
extended Chaplygin cosmologies with a big rip. {\em Phys. Rev. D} {\bf 
2006}, 74, 044022, DOI: 10.1088/0264-9381/23/9/027

\bibitem{etc12} Cai, Y.-F., Li, H., Piao, Y.-S., Zhang, X. Duality 
invariance and cosmological dynamics.  {\em Phys. Lett. B} {\bf 2007}, 
646, 141, DOI: 10.1142/S0218271808013820

\bibitem{etc13} Chimento, L.P., Devecchi, F.P., Forte, M.I., Kremer, G.M. 
Phantom cosmologies and fermions. {\em Class. Quantum Grav.} {\bf 2008}, 
25, 085007, DOI: 10.1088/0264-9381/25/8/085007

\bibitem{etc14} Cataldo, M., Chimento, L.P., Form invariant 
transformations between n-dimensional flat Friedmann-Robertson-Walker 
cosmologies.  {\em Int. J. Mod. Phys. D} {\bf 2008}, 17, 1981-1989, 
https://doi.org/10.1142/S0218271808013790

\bibitem{etc15} Capozziello, S., Piedipalumbo, E., Rubano, C., Scudellaro, 
P., Noether symmetry approach in phantom quintessence cosmology. {\em 
Phys. Rev. D} {\bf 2009}, 80, 104030, DOI: 10.1103/PhysRevD.80.104030

\bibitem{etc16} Wang, J., Lan, T., Yang, S.-P., Cosmic Duality and 
Statefinder Diagnosis of Spinor Quintom. {\em J. Theor. Phys.} {\bf 2012}, 
1, 62-75.

\bibitem{etc17} Capozziello, S., Faraoni, V., {\em Beyond Einstein 
Gravity}; Springer: New York, 2010.

\bibitem{etc18} Cai, Y.-F., Saridakis, E.N., Setare, M.R., Xia, J.-Q.  
Quintom Cosmology: Theoretical implications and observations. {\em Phys. 
Rep.} {\bf 2010}, 493, 1-60, DOI: 10.1016/j.physrep.2010.04.001

\bibitem{etc19} Pucheu, L., Bellini, M. Phantom and inflation scenarios 
from a 5D vacuum through form-invariance transformations of the Einstein 
equations. {\em Nuovo Cimento B} {\bf 2010}, 125, 851-859, DOI: 
10.1393/ncb/i2010-10888-0

\bibitem{etc20} Chimento, L.P., Lazkoz, R., Richarte, M.G. Inflation in 
the Dirac-Born-Infeld framework. {\em Phys. Rev. D} {\bf 2011}, 83, 
063505, DOI: 10.1103/PhysRevD.83.063505

\bibitem{myPLB} Faraoni, V. A symmetry of the spatially flat Friedmann 
equations with barotropic fluids. {\em Phys. Lett. B} {\bf 2011}, 703, 
228–231, https://doi.org/10.1016/j.physletb.2011.08.018

\bibitem{Wald} Wald, R.M.  {\em General Relativity}; Chicago University 
Press: Chicago, IL, USA, 1984.

\bibitem{Carroll} Carroll, S.M. {\em Spacetime and Geometry: An 
Introduction to General Relativity};  Addison Wesley: San Francisco, 2004.

\bibitem{etc21} Rosu, H.C., Khmelnytskaya, K.V. Shifted Riccati procedure: 
Application to conformal barotropic FRW cosmologies. {\em SIGMA} {\bf 
2011}, 7, 013, DOI: 10.3842/SIGMA.2011.013

 \bibitem{etc22} Rosu, H.C., Ojeda-May, P. Supersymmetry of FRW barotropic 
cosmologies. {\em Int. J. Theor. Phys.} {\bf 2006}, 45, 1191, DOI: 
10.1007/s10773-006-9123-2

\bibitem{etc23} Nowakoswki, M., Rosu, H.C. Newton`s Laws of motion in the 
form of a Riccati equation. {\em Phys. Rev. E} {\bf 2002}, 65, 047602, 
DOI: 10.1103/PhysRevE.65.047602

\bibitem{Faraoniunpublished} Faraoni, V. {\bf 2020}, Analogy 
between freezing lakes and the cosmic radiation era (unpublished).

\end{thebibliography}}
\end{document}